\documentclass[10pt,a4paper,prd,reprint,superscriptaddress]{revtex4-1}
\usepackage[T1]{fontenc} \usepackage{pstricks} \usepackage{subfigure}
\usepackage[normalem]{ulem}
\usepackage{graphicx} 
\usepackage{bm}       
\usepackage{amsbsy}
\usepackage{amsfonts}
\usepackage{amsmath}
\usepackage{amssymb}
\usepackage{fixmath}
\usepackage[colorlinks]{hyperref}
\usepackage{physics}
\usepackage{color}
\hypersetup{
	bookmarksnumbered,
	pdfstartview={FitH},
	citecolor={darkgreen},
	linkcolor={blue},
	urlcolor={darkblue},
	pdfpagemode={UseOutlines}}
\definecolor{darkgreen}{RGB}{20,100,20}
\definecolor{darkblue}{RGB}{0,0,130}
\definecolor{darkred}{rgb}{.8,0,0}

\renewcommand{\vec}[1]{\mathbold{#1}}

\begin{document}	
	\title{Measuring the Casimir-Polder potential with Unruh-DeWitt detector excitations}
	\author{Kacper D\k{e}bski}
	\email{k.debski@student.uw.edu.pl}
	\affiliation{Institute of Theoretical Physics, University of Warsaw, Pasteura 5, 02-093 Warsaw, Poland}
	\author{Piotr T. Grochowski}
	\email{piotr@cft.edu.pl}
	\affiliation{Center for Theoretical Physics, Polish Academy of Sciences, Aleja Lotnik\'ow 32/46, 02-668 Warsaw, Poland}
	\author{Andrzej Dragan}
	\email{dragan@fuw.edu.pl}
	\affiliation{Institute of Theoretical Physics, University of Warsaw, Pasteura 5, 02-093 Warsaw, Poland}
	\affiliation{Centre for Quantum Technologies, National University of Singapore, 3 Science Drive 2, 117543 Singapore, Singapore}
	\date{\today}

	\begin{abstract}
		We propose a new method to measure Casimir-Polder force in an optical cavity by means of atomic excitations.
		We utilize a framework in which Unruh-DeWitt Hamiltonian mimics the full matter-field interaction and find a connection between the excitation rate of the two-level atom and the Casimir-Polder energy, allowing to map one onto the other.
		We argue that such a realization opens a route to study Casimir-Polder potentials through measurement of excited state's population in large, spatially compact ensembles of atoms.
	\end{abstract}
	\maketitle
	\section{Introduction}
	The idea of light quantization is one of the cornerstones of the quantum theory.
	Among many of its proofs, experiments involving a weak electromagnetic field played a crucial role in establishing the consensus.
	The seminal paper of Casimir and Polder shows that even in the limit of a photonless state -- the quantum vacuum -- a neutral atom near a dielectric wall feels a force mediated through the quantum fluctuations~\cite{Casimir1948}.
	Not before 1990s was such a behavior experimentally demonstrated, providing another confirmation of the quantization of the light and opening perspectives on investigating the quantum vacuum~\cite{Sukenik1993,Lamoreaux1997,Mohideen1998}.
	
	The presence of Casimir-Polder forces was demonstrated in various setups, involving different types of plates, from conducting to dielectric ones, and different types of probes, from atomic, through mechanical, to Bose-Einstein condensates~\cite{Schwinger1978, Balian1978, Plunien1986,Bordag2001,Harber2005a,Obrecht2007,Klimchitskaya2009}.
	In most cases, Casimir forces are attractive, however it was proposed that a repulsive character of the interaction is also possible~\cite{Dzyaloshinskii1961,Boyer1974,Bordag2001,Kenneth2002,Levin2010}, as recently showed experimentally~\cite{Munday2009}.
	
	The QED minimal coupling is usually a model of choice in describing Casimir phenomena, however there are others, providing an effective description of the electromagnetic field.
	One of the minimal models allowing system to exhibit both attractive and repulsive character of Casimir force is the Unruh-DeWitt (UDW) model~\cite{DeWitt1979} in the optical cavity~\cite{Alhambra2014}.
	
	Routinely used as a pointlike particle detector in the quantum field theory~\cite{Crispino2008,Birrell1982,Louko2008,Hodgkinson2013,Brown2013}, UDW model constitutes of a two-level atom coupled to the scalar field.
	Despite being noticeably simpler than the full QED Hamiltonian, it was shown to be a reasonable approximation to the atom-field interaction when no orbital angular momentum is exchanged~\cite{Martin-Martinez2013a}.
	Moreover, it was used to study Casimir-Polder forces~\cite{Passante2003, Rizzuto2004,Spagnolo2006,Rizzuto2007}.
	
	In this work, we analyze UDW model in the optical cavity, focusing on the following two aspects.
	Using the second order perturbation theory, we calculate the Casimir-Polder (CP) potential as a function of the atom's position in the cavity.
	We find that under some assumptions, it is intrinsically connected to the excitation probability of the two-level system that models the atom.
	As a consequence, we propose a simple experimental scenario of measuring the CP force by means of the atomic excitations in a Bose-Einstein condensate placed near the wall.
	
	The work is structured as follows.
	In Sec. II we introduce the model, underline its connection to the full QED interaction Hamiltonian, and compute both CP potential and the excitation rate of the UDW detector.
	In Sec. III we find the connection between the two and present the toy experimental proposal.
	The final Sec. IV concludes the manuscript with the recapitulation and the outlook.
	
	\section{Model}
	We will work in natural units, $\hbar=c=e=1$.
	Let us consider a scalar field of a mass $m$ governed by the Klein-Gordon equation:
	\begin{equation}
	\left(\Box+m^2  \right) \hat{\phi}=0
	\end{equation}
	in the cavity with a length of $L$ fulfilling Dirichlet boundary conditions, $\hat{\phi}(x=0)=\hat{\phi}(x=L)=0$
	with the following mode solutions:
	\begin{equation}
	u_n(x,t)=\frac{1}{\sqrt{\omega_n L}} \sin{\left(k_n x\right)}e^{-i\omega_n t}\equiv u_n(x)e^{-i\omega_n t},    
	\end{equation}
	where $\omega_n=\sqrt{k_n^2+m^2}$, $k_n=\frac{n\pi}{L}$, $n \in \mathbb{Z}$.
	Using these modes, the field $\hat{\phi}$ can be decomposed as:
	\begin{equation}
	\hat{\phi}(x)=\sum_{n}\left[\hat{a}_n^{\dagger} u_{n}^*(x)+\hat{a}_n u_n(x)\right],
	\end{equation}
	where $\hat{a}_n$ and $\hat{a}_n^{\dagger}$ are annihilation and creation bosonic operators satisfying the canonical commutation relations, $\left[\hat{a}_n,\hat{a}_k^{\dagger}\right]=\delta_{nk}$ and $\left[\hat{a}_n,\hat{a}_k\right]=\left[\hat{a}_n^{\dagger},\hat{a}_k^{\dagger}\right]=0$.
	
	In the distance $d$ from the boundary let us place a two-level system corresponding to the simplest model of an atom with an energy gap $\Omega$.
	Such a description approximates e.g. a hydrogen atom which is not placed in a strong classical background field and therefore no transition is resonantly coupled to the cavity.
	As we are interested in working with the vacuum state of the cavity, this proves to be a valid approximation.
	
	Then, the full Hamiltonian of the considered model includes free Hamiltonians of the scalar field and of the atom, and the term accounting for the interaction between both of them, $\hat{H}_{\mathrm{I}}$. 
	One of the simplest possible choices of the interaction between the scalar field and the two-level system is the pointlike Unruh-DeWitt Hamiltonian.
	In the Schr\"odinger picture it takes the following form:
	\begin{equation}
	\hat{H}_{\mathrm{UDW}}=\lambda~\hat{\mu}_{\mathrm{S}}~\hat{\phi}(x),
	\label{eqn:UdWhamiltonian}
	\end{equation}
	where $\lambda$ -- dimensionless coupling constant, $\hat{\mu}_{\mathrm{S}}$ -- the monopole moment of the detector, $\hat{\mu}_{\mathrm{S}}=\hat{\sigma}^{+}+\hat{\sigma}^{-}=\ket{g}\bra{e}+\ket{e}\bra{g}$, where $\ket{g}$ is the ground state of the two-level system and $\ket{e}$ is its excited state.
	Moreover, $\hat{\phi}(x)$ is the scalar field operator evaluated at the point at which the pointlike detector is placed.
	In the spirit of electromagnetic field considerations, this first order term would be called a paramagnetic one.
	
	This simple model of interaction can be extended to a more realistic form including the second order term in the Hamiltonian corresponding to a diamagnetic, self-interaction term of the full QED Hamiltonian.
	This quadratic term of the Unruh-DeWitt Hamiltonian has the form:
	\begin{align}
	\hat{H}_{\mathrm{UDW}}^2&=\left(\lambda\left(\ket{g}\bra{e}+\ket{e}\bra{g}\right)\hat{\phi}\right)^2 \nonumber \\
	&=\lambda^2\left(\ket{g}\bra{g}+\ket{e}\bra{e}\right)\hat{\phi}^2=\lambda^2\hat{\phi}^2.
	\end{align}
	It is worth to notice that such a term does not change a detector state.
	
	At this point, let us revoke some key results from Refs.~\cite{Martin-Martinez2013a} that compare QED Hamiltonian and UDW one.
	The minimal electromagnetic coupling in the Coulomb gauge reads:
	\begin{equation}
	\hat{H}_{\mathrm{QED}} = -\frac{1}{m} \vec{A}(\vec{x}) \cdot \vec{p}+\frac{1}{2 m} \left[ \vec{A}(\vec{x})\right]^2. 
	\end{equation}
	The main difference is the vector character of the EM interaction in contrast to the scalar one that we consider.
	However, a scalar field can be readily utilized to describe electric and magnetic contributions separately, given appropriate boundary conditions.
	Indeed, such a description has been used to analyze Casimir-Polder interaction in the past.
	It has to be noted that such a scalar model does not allow any exchange of the orbital momentum, however we retreat to the simple case of atomic transitions that obey this rule.
	
	As mentioned above, the QED Hamiltonian consists of two terms -- paramagnetic and diamagnetic ones.
	The simplified light-matter interaction Hamiltonians often neglect the second term while working with weak fields.
	The minimal coupling in the vacuum implies interaction only with the quantum fluctuations $<\vec{A}^2>$.
	However, in the vacuum, the value of $<\vec{A}^2>$ depends on the region in which an atom resides -- or in the language of quantum field theory -- on the region these quantum fluctuations are smeared over.
	It happens that while approaching a limit of a pointlike atom (detector), this variance diverges.
	So, it introduces a necessity to allow for a finite size of the atom, unlike in the simple UDW model.
	
	In the original Casimir and Polder paper, such a problem was also present -- it was taken care of by the means of introducing a regularizing factor $e^{-\gamma k}$ in the integrals over momentum space.
	However, we follow a procedure used by \cite{Martin-Martinez2013a}, where an explicit spatial form of the ground state of the atom is assumed:
	\begin{equation}\label{pro}
	\Psi(x)=\frac{e^{-x/a_0}}{a_0},
	\end{equation}
	where $a_0$ is some characteristic length associated to the spherically symmetric atomic profile (meant to be of the order of magnitude of Bohr radius).
	Such an approach modifies the UDW Hamiltonian by effectively coupling the detector to an effective field,
	\begin{equation}
	\hat{\phi}_{\mathrm{R}}(x)=\sum_{n}f_n\left[\hat{a}_n^{\dagger} u_{n}^*(x)+\hat{a}_n u_n(x)\right],
	\end{equation}
	where
	\begin{equation}f_n=\frac{2}{\left(a_0 k_n\right)^2+1}
	\end{equation}
	are Fourier transforms of the spatial profile \eqref{pro} evaluated at momentum $k_n$.
	Such a momentum-space profile is typical for zero angular momentum orbitals and can describe the simplest case of a hydrogen atom and its lowest transition, 1s $\rightarrow$ 2s.
	
	The next simplification of the UDW model involves assuming equal contributions from both of the nondiagonal parts of the Hamiltonian acting on the space spanned by the internal states of the atom.
	In a general case, their relative weight can be unequal, but in the case of a spherical symmetry of both the ground and the excited states, they happen to be equal.
	
	The other difference between QED and UDW Hamiltonians come from the fact that in the former the relative strength of para- and diamagnetic terms is given explicitly.
	It is not the case in the latter, as it has to be computed for specific profiles of the ground and the excited states.
	It can be done, however we will take the advantage of our model by considering a general, dimensionless parameter quantifying this relative strength.
	
	Combining all of these considerations, we finally get the extended version of the UDW Hamiltonian that mimics the QED one:
	\begin{equation}\label{ham}
	\hat{H}_{\mathrm{I}}=\lambda\left(\ket{g}\bra{e}+\ket{e}\bra{g}\right)\hat{\phi}_{\mathrm{R}}+\alpha\frac{\lambda^2}{\Omega}\hat{\phi}_{\mathrm{R}}^2,
	\end{equation}
	where $\alpha$ is a dimensionless constant and a free parameter to tune the relative strength between para- and diamagnetic terms.
	Energy $\Omega$ is introduced here to provide the correct units.
	Such a Hamiltonian can effectively mimic some forms of the full QED interaction~\cite{Martin-Martinez2013a}.
	It has to be noted that such a Hamiltonian is only a one-dimensional toy model that is utilized to model the qualitative effects coming from the full electromagnetic one.
	By keeping free parameters $\lambda$ and $\alpha$ explicitly in the calculations, we will show that some interesting conclusions stay the same for their arbitrary values.
	
	\subsection{Casimir-Polder potential}
	The first step is to find how the full energy of the system changes with the position of the atom in the cavity.
	The difference between this full energy, $E$, and the sum of the ground state energies of noninteracting cavity and the atom, $E_0$, is called the Casimir-Polder potential, $E_{\mathrm{CP}}$.
	The usual Casimir-Polder force acting on the atom in the fixed cavity is then understood as a spatial derivative of the Casimir-Polder potential, $F=-\nabla E_{\mathrm{CP}}$.
	We consider a system prepared in the state $\ket{g,0}$ -- the scalar field is in the vacuum state and two-level system is in the ground state.
	The system is then slightly perturbed by the extended UDW Hamiltonian~\eqref{ham} with $\lambda$ being the perturbation parameter.
	We will calculate the following energy in the second order of the perturbation theory.
	It takes form:
	\begin{eqnarray}
	E&=&E_0+E^{(1)}+E^{(2)}+\mathcal{O}(\lambda^4), \nonumber \\ 
	E^{(1)}&=&\bra{g,0}\hat{H}_{\mathrm{I}}\ket{g,0}, \nonumber \\
	E^{(2)}&=&\sum_{n=0}^{\infty}\sum_{s=\{g,e\}}\frac{|\bra{s,n}\hat{H}_{\mathrm{I}}\ket{g,0}|^2}{E_0-(E_0+\omega_n+\Omega_s)},
	\end{eqnarray}
	where state $\ket{s,n}$, $s\in\{g,e\}$ corresponds to the arbitrary final state of the atom and the scalar field in the state $\ket{n}$ of energy $\omega_n$.
	Furthermore, $\Omega_s$ is the energy of the detector in the state $\ket{s}$, meaning that $\Omega_g=0$ and $\Omega_e=\Omega$.
	Then, we have:
	\begin{eqnarray}
	E^{(1)}&=&\frac{\alpha\lambda^2}{\Omega}\bra{0}\hat{\phi}_{\mathrm{R}}^2\ket{0}=\frac{\alpha\lambda^2}{\Omega L}\sum_{n=1}^{\infty}\frac{f_n^2 \sin^2{\left(k_n x\right)}}{\omega_n},\nonumber\\
	E^{(2)}&=&-\sum_{n=1}^{\infty}\frac{\lambda^2}{\omega_n+\Omega}|\bra{n}\hat{\phi}_{\mathrm{R}}\ket{0}|^2+\mathcal{O}(\lambda^4)\nonumber\\
	&=&-\sum_{n=1}^{\infty}\frac{\lambda^2}{\omega_n+\Omega}\frac{f_n^2 \sin^2{\left(k_n x\right)}}{\omega_n L}+\mathcal{O}(\lambda^4).\nonumber
	\end{eqnarray}
	It is useful to note that the term $f_n$ makes $E^{(1)}$ convergent.
	The whole second-order Casimir-Polder potential then reads
	\begin{align}
	E_{\mathrm{CP}}&=E^{(1)}+E^{(2)} \nonumber \\
	&=\lambda^2\sum_{n=1}^{\infty}\frac{f_n^2\sin^2{\left(k_n x\right)}}{\omega_n L \left(\omega_n+\Omega\right)}\left[\left(\alpha-1\right)+\alpha \frac{\omega_n}{\Omega}\right].
	\label{eqn:ECP}
	\end{align}
	One can immediately see that depending on the parameter $\alpha$, the Casimir-Polder potential, and consequently Casimir-Polder force can be either positive or negative.
	It confirms the usual phenomenology in which Casimir forces can be either repulsive or attractive, depending on the physical scenario involved.
	
	\subsection{Probability of excitation}
	The next step is to assume the same physical model but now with the interaction lasting for some finite time $\sigma$.
	As the electromagnetic interaction cannot be switched on or off, we choose to interpret the finite interaction time as a time between the creation of a setup and a destructive measurement.
	We aim to find the probability of measurement of the excited state of the atom, as it was initially prepared unexcited in the cavity.
	Therefore, we have to define the time-dependent Hamiltonian of interaction, allowing for a finite time interaction.
	We can modify previously showed model by adding a time-dependent switching function $\chi(t)$.
	The modified, time-dependent version of the extended UDW Hamiltonian in the Schr\"odinger picture has the following form:
	\begin{equation}
	\hat{H}_{\mathrm{UDW}}(t) 
	=\chi(t)\left[\lambda~\hat{\mu}_{S} (t)~\hat{\phi}_{\mathrm{R}}(x)
	+
	\alpha \frac{\lambda^2}{\Omega}\left(\hat{\phi}_{\mathrm{R}}(x)\right)^2
	\right].
	\end{equation}
	We assume that the interaction starts and ends rapidly, so that $\chi(t)=1$ for $t\in(0,\sigma)$ and $\chi(t)=0$ for any other time.
	As it was mentioned before, the full Hamiltonian includes also a time-independent free scalar field and a free two-level system part.
	We proceed to use the Dirac picture, because the full Hamiltonian contains time-independent $\hat{H}_0=\sum_n \omega_n \hat{a}_n^{\dagger}\hat{a}_n\otimes\Omega\hat{\sigma}^{+}\hat{\sigma}^{-}$ and a time-dependent interaction component %$\hat{H}_{\mathrm{I}}(t)$
	coming from the Unruh-DeWitt interaction.
	The evolution in such a scenario is given by operator in the form: $\hat{U}
	=\mathcal{T}\exp{
		-i\int_{-\infty}^{\infty} 
		\mathrm{d}t\hat{H}_{\mathrm{I}}^{(D)}(t)
	}$, where $(D)$ represents operator in the Dirac picture.
	
	As a result of the interaction, the state of the field can be changed, however we are interested only in finding the probability of the detector's excitation.
	The final state after the interaction between the detector and the scalar field can be written as $\ket{e,l}$, where $l\in\mathbb{N}\cup\{0\}$.
	Using the Born rule, we can write probability of excitation $p_{g\xrightarrow{}e}$ in the form:
	\begin{equation}
	p_{g\xrightarrow{}e}=\sum_{l\in\mathbb{N}\cup\{0\}}|\bra{e,l}
	\int_{-\infty}^{\infty} \mathrm{d}t
	\hat{H}_{\mathrm{I}}^{(D)}(t)         \ket{g,0}|^2.
	\label{eqn:p2}
	\end{equation}
	
	The extended UDW Hamiltonian in the Dirac representation reads:
	\begin{eqnarray} \label{ham2}
	\hat{H}_{\mathrm{I}}^{(D)}(t) 
	=\chi(t)\left[\lambda~\hat{\mu}^{(D)}~\hat{\phi}_{\mathrm{R}}^{(D)}
	+
	\frac{\alpha\lambda^2}{\Omega}\left(\hat{\phi}_{\mathrm{R}}^{(D)}\right)^2
	\right],
	\end{eqnarray}
	where:
	\begin{eqnarray}
	\hat{\mu}^{(D)}&=&\left(e^{i\Omega t} \hat{\sigma}^{+}+e^{-i\Omega t} \hat{\sigma}^{-}\right),\\
	\hat{\phi}_{\mathrm{R}}^{(D)}(x)&=&\sum_{n}f_n\left[\hat{a}_n^{\dagger} u_{n}(x)e^{i\omega_n t}+H.c.\right].
	\end{eqnarray}
	
	Only the first part, linear in the coupling constant $\lambda$ contains an operator changing the state of the detector.
	The second-order term does not contribute to the probability of excitation given by the equation 
	\eqref{eqn:p2}, because $\bra{e} \frac{\alpha\lambda^2}{\Omega}\left(\hat{\phi}_{\mathrm{R}}^{(D)}\right)^2\ket{g}=0$.
	After some direct calculation, by plugging~\eqref{ham2} in~\eqref{eqn:p2}, we get:
	\begin{equation}
	p_{g\xrightarrow{}e}=4\lambda^2\sum_{n=1}^{\infty}\frac{f_n^2\sin^2{\left(k_n x\right)}}{\omega_n L}\frac{\sin^2{\left[\frac{1}{2}\sigma\left(\omega_n+\Omega\right)\right]}}{\left(\omega_n+\Omega\right)^2}.
	\label{eqn:skonczoneprawd}
	\end{equation}
	
	The above result corresponds to an arbitrarily chosen time of interaction $\sigma$, but if the measurement apparatus does not have a time resolution good enough to work withing the scale of an atomic transition $1/\Omega$ one has to consider a coarse grained version of the above formula. 
	Such an averaged out excitation probability reads:
	\begin{align}
	p_{g\xrightarrow{}e}^{\text{av}}&=4\lambda^2\int \mathrm{d}\sigma \sum_{n=1}^{\infty}\frac{f_n^2\sin^2{\left(k_n x\right)}}{\omega_n L}\frac{\sin^2{\left[\frac{1}{2}\sigma\left(\omega_n+\Omega\right)\right]}}{\left(\omega_n+\Omega\right)^2} \nonumber \\
	&=
	2\lambda^2 \sum_{n=1}^{\infty}\frac{f_n^2\sin^2{\left(k_n x\right)}}{\omega_n L \left(\omega_n+\Omega\right)^2}
	.
	\label{eqn:pav} 
	\end{align}
	
	This new quantity is no longer dependent on the interaction time and is an explicit function of a distance from the wall.
	In the next Section we will show the connection between this averaged probability and the Casimir-Polder force.
	
	\section{Retrieving Casimir-Polder potential from the average excitation rate}
\begin{figure}[h!tbp]
		\includegraphics[width=1\linewidth]{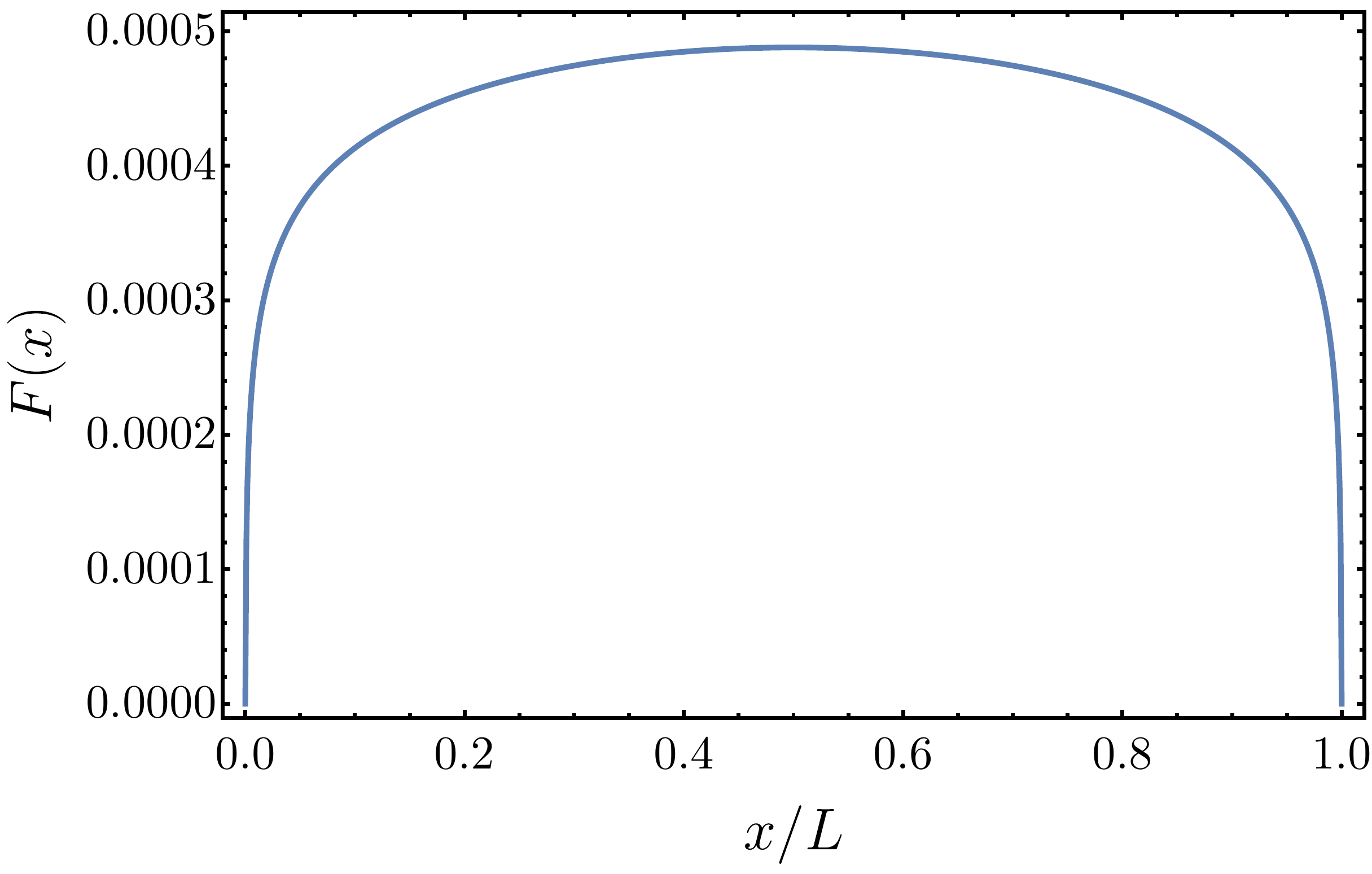}
		\includegraphics[width=1\linewidth]{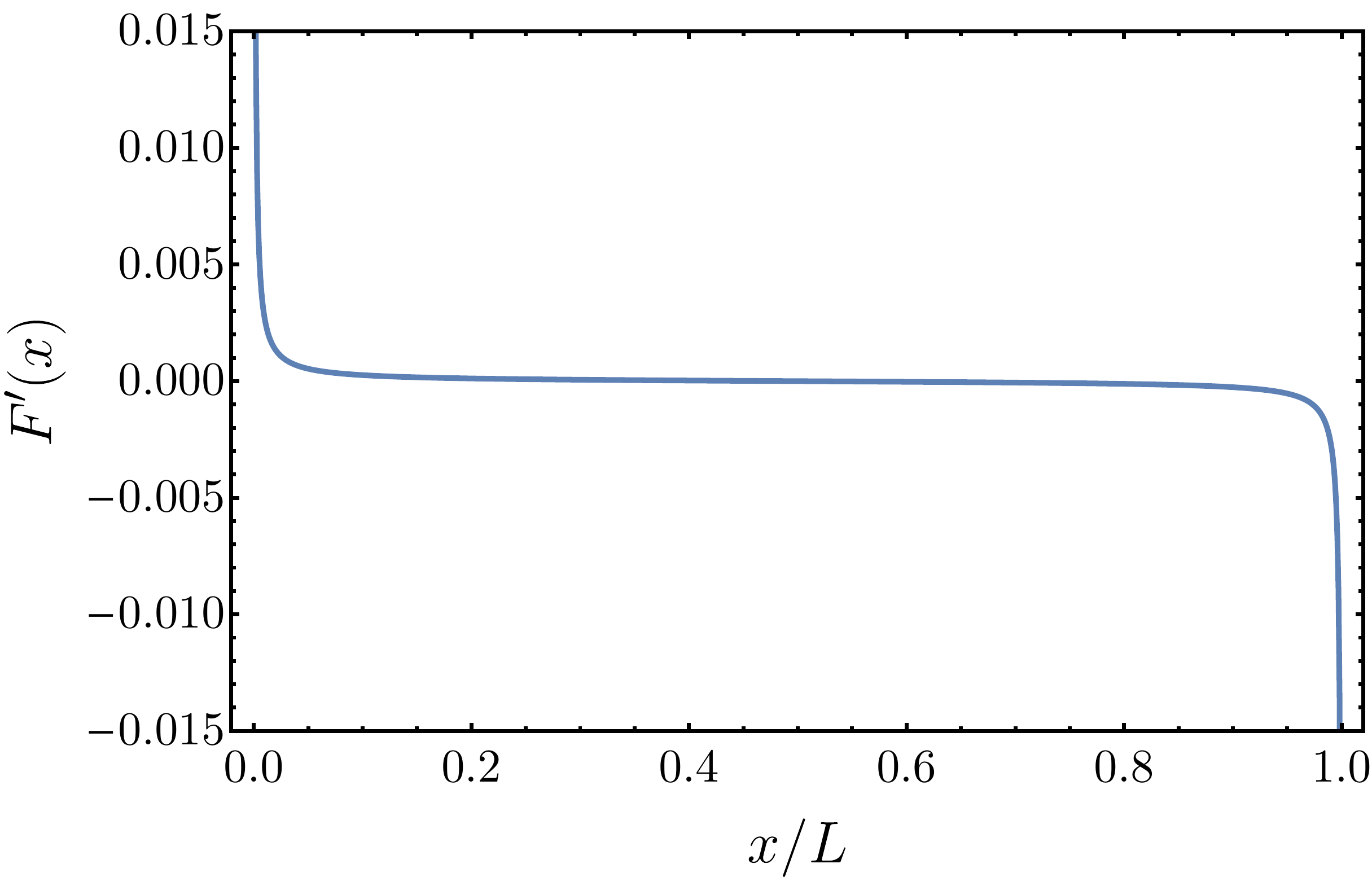}
		\caption{(Top) Universal function $F(x)$ proportional to both Casimir-Polder potential and the average excitation probability for the atom at position $x/L$ in the cavity.
			The parameters taken for the plot read: $L=1$, $m=1 \cdot 10^{-3}$, $\lambda =1 \cdot 10^{-2} $, $\Omega=1$.
			(Bottom) The derivative of the universal function $F(x)$, corresponding to the Casimir-Polder force in the optical cavity.
			\label{fig1}}
	\end{figure}

	We now proceed to compare both results.
	Both energy and probability given by Eqs. \eqref{eqn:ECP} and \eqref{eqn:pav} are represented by infinite series.
	We find that contributing terms in both of these series occur only for small (in comparison to the energy gap $\Omega$) values of $n$ , $\omega_n \ll \Omega$.
	For a numerical analysis of that fact, see Appendix. 
	Therefore, we can treat $\frac{\omega_n}{\Omega}$ as a small parameter and expand both \eqref{eqn:ECP} and \eqref{eqn:pav} up to the first subleading order:
	\begin{equation}
	\label{cpeneeny}
	E_{\mathrm{CP}}\approx \Omega \sum_{n} p_n (x) \left[ \left( \alpha-1\right)+\frac{\omega_n}{\Omega}  \right] 
	\end{equation}
	and 
	\begin{equation}
	p_{g\xrightarrow{}e}^{\text{av}}\approx 2 \sum_{n} p_n (x) \left[1-2\frac{\omega_n}{\Omega}  \right] 
	\end{equation}
	where
	\begin{equation}
	p_n (x)  = \frac{\lambda^2 f_n^2 \sin^2\left( k_n x \right) }{\omega_n L \Omega^2}.
	\end{equation}
	The universal function
	\begin{equation}
	F(x)=\sum_{n} p_n (x)
	\end{equation}
	that reproduces the general shape of both Casimir-Polder potential and the averaged excitation probability is shown at Fig.~\ref{fig1} together with its derivative, corresponding to the Casimir-Polder force.
	
	Up to the leading order in $\omega_n/\Omega$, we have a proportionality between the Casimir-Polder energy and the averaged excitation probability of the atom:
	\begin{equation} \label{res}
	E_{\mathrm{CP}} \approx \frac{1}{2} \Omega \left(\alpha - 1 \right) p_{g\xrightarrow{}e}^{\text{av}}.
	\end{equation}
	The proportionality constant is a function of the energy gap and the internal properties of the atom, implicitly contained in the parameter $\alpha$.
	Unless $\alpha$ is extremely fine tuned to be 1, the proportionality between $E_{\mathrm{CP}}$ and $p_{g\xrightarrow{}e}^{\text{av}}$ is preserved.
	One has to note that here is no realistic value of $\alpha$ for the three-dimensional electromagnetic model of interaction between atom an the field, as the presented one-dimensional toy model is aimed only at grasping qualitative effects present in the system.
	However, within this toy model, taking $a_0=10^{-2}$ in natural units, $\alpha$ can be calculated to be $\alpha \sim 1/400$~\cite{Alhambra2014a}.
	It shows that at least for hydrogen-like atoms in 1D toy model, value $\alpha \sim 1$ is not a typical one.
	
	The result~\eqref{res} shows that Casimir-Polder potential can be indirectly recovered through analyzing the rate of atomic excitation near the wall. 
	However, the probability of exciting a single atom through the interaction with the vacuum fluctuations is extremely small.
	It can be in some way adjusted by a proper choice of atomic species and of a proper atomic transition, however the most straightforward way is to use a large number of atoms that are placed within the same region.
	
	The natural candidate for such an ensemble is a Bose-Einstein condensate placed near the wall of the optical cavity. Modern ultracold experiments provide a clean environment to probe such settings.
	The first perk is the localization of the atoms by the means of optical trapping -- an ultracold sample can be placed in highly localized region of space (given by the usual Thomas-Fermi radius of BEC).
	The second one is a fine control over the parameters of neutral atoms involved -- the size of the atomic cloud is tuned with the trapping and with the two-body interaction between the atoms, that can be effectively turned off by the means of Fesbach resonances.
	
	Indeed, Bose-Einstein condensates have already been utilized to probe Casimir forces through their effect on the collective modes excited in the atomic cloud.
	In Ref.~\cite{Obrecht2007} a BEC of rubidium was placed near the dielectric surface and the frequency of dipole mode was probed.
	It allowed to measure nonzero temperature Casimir force acting on individual atoms.
	We, however, propose a different scheme in which the Casimir force is not directly measured through the change of the collective motion, but rather the potential itself is analyzed through the excitation of atoms.
	
	First, the BEC is treated as a collection of a large number of individual atoms that is spatially tightly packed and well described by the Thomas-Fermi profile and not as interacting many-body system.
	The hydrogen-like atom assumption we have is a shortcoming, however condensates of such atoms have been produced.
	We stress that our aim is not to perform the full three-dimensional calculation of the realistic QED system, but rather to show that at the level of a toy model mimicking the QED, there exists a direct relation between the Casimir-Polder energy and the excitation probability of an atom involved.
	It is worth checking how the averaging over the typical density of BEC changes this relation.
	
\begin{figure}[h!tbp]
		\includegraphics[width=1\linewidth]{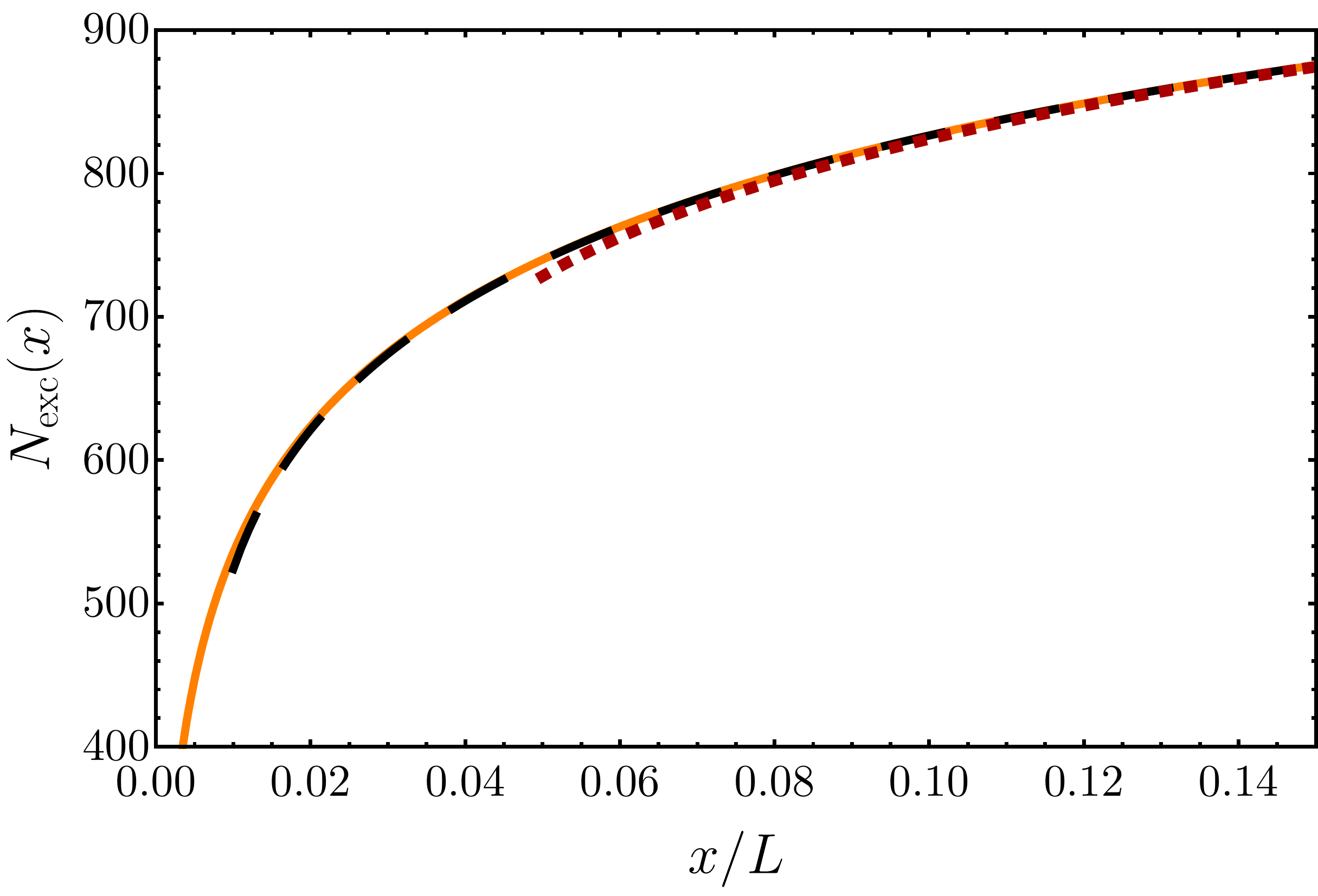}
		\caption{Comparison between time-averaged excitation probability averaged over the density of atomic Bose-Einstein condensate for different Thomas-Fermi radii (red, dotted line -- $R_{TF} =0.05$, black, dashed line  -- $R_{TF} =0.01$, orange, solid line -- $R_{TF} \rightarrow 0$).
			The spatial averaging does not introduce appreciable quantitative difference, yielding same curve for each radius, corresponding to the Casimir-Polder potential.
			Note that each curve starts at $R_{TF}$ due to the finite size of the BEC used as a probe.
			The parameters taken for the plot read: $L=1$, $m=1 \cdot 10^{-3}$, $\lambda =1 \cdot 10^{-2} $, $\Omega=1$.
			\label{fig2}}
	\end{figure}

	The Thomas-Fermi single-particle density of BEC with perpendicular degrees of freedom integrated out and situated at $x_0$ from the boundary read:
	\begin{equation} \label{bec}
	n(x;x_0)=\frac{15 N}{16 R_{TF}} \left[1-\left(\frac{x-x_0}{R_{TF}} \right)^2  \right]^2, 
	\end{equation}
	where $N$ -- total number of atoms in the condensate and $R_{TF}$ -- Thomas-Fermi radius.
	If we follow the excitation scheme in which multiple single-shot experiments measuring population of the excited state (e.g. via in situ imaging) are performed, the averaged population of excited state takes the form:
	\begin{equation} \label{ave}
	N_{\text{exc}} (x)=\int p_{g\xrightarrow{}e}^{\text{av}}(u) n(u;x)  \dd u.
	\end{equation} 
	The effect of such an averaging, for different sizes of BECs is shown at Fig.~\ref{fig2}.
	It appears that averaging over the density of BEC does not introduce appreciable changes to the spatial dependence of population of excited states in comparison to the Casimir-Polder potential. 
	
	\section{Recapitulation and outlook}
	To summarize, we have analyzed a system consisting of an atom (described by a two-dimensional Hilbert space) interacting with a scalar field within a one-dimensional cavity.
	We have argued that an extended version of the Unruh-DeWitt Hamiltonian coupled to the scalar Klein-Gordon field provides a qualitatively reasonable approximation to the full light-matter interaction when the vacuum state of the cavity is involved.
	Utilizing the second-order perturbation theory, we have calculated the Casimir-Polder energy of the system and excitation probability of the atom when placed in a fixed distance from the wall of the cavity.
	We have shown that up to the leading order, both of these quantities coincide with each other up to multiplicative constant depending on the internal structure of the atom.
	As a result, we have argued that a suitable experimental setup involving measurement of the population of the excited atomic state in e.g. ultracold system of a two-level Bose-Einstein condensate, can be used to measure Casimir-Polder phenomena.
	Moreover, we have checked how averaging over the BEC density affects the above-mentioned relation between CP energy and the excitation probability.
	
	As a future line of work, a natural consequence would be a consideration a full three-dimensional system in a realistic experimental scenario.
	Another potential research could involve studying applicability of the Untuh-DeWitt Hamiltonian in describing neutral atoms in optical cavities.
	
	\section{Acknowledgments}
	P.T.G. appreciates discussions with K. Rz\k{a}\.{z}ewski.

%	\bibliography{Casimir}

\begin{thebibliography}{29}%
	\makeatletter
	\providecommand \@ifxundefined [1]{%
		\@ifx{#1\undefined}
	}%
	\providecommand \@ifnum [1]{%
		\ifnum #1\expandafter \@firstoftwo
		\else \expandafter \@secondoftwo
		\fi
	}%
	\providecommand \@ifx [1]{%
		\ifx #1\expandafter \@firstoftwo
		\else \expandafter \@secondoftwo
		\fi
	}%
	\providecommand \natexlab [1]{#1}%
	\providecommand \enquote  [1]{``#1''}%
	\providecommand \bibnamefont  [1]{#1}%
	\providecommand \bibfnamefont [1]{#1}%
	\providecommand \citenamefont [1]{#1}%
	\providecommand \href@noop [0]{\@secondoftwo}%
	\providecommand \href [0]{\begingroup \@sanitize@url \@href}%
	\providecommand \@href[1]{\@@startlink{#1}\@@href}%
	\providecommand \@@href[1]{\endgroup#1\@@endlink}%
	\providecommand \@sanitize@url [0]{\catcode `\\12\catcode `\$12\catcode
		`\&12\catcode `\#12\catcode `\^12\catcode `\_12\catcode `\%12\relax}%
	\providecommand \@@startlink[1]{}%
	\providecommand \@@endlink[0]{}%
	\providecommand \url  [0]{\begingroup\@sanitize@url \@url }%
	\providecommand \@url [1]{\endgroup\@href {#1}{\urlprefix }}%
	\providecommand \urlprefix  [0]{URL }%
	\providecommand \Eprint [0]{\href }%
	\providecommand \doibase [0]{http://dx.doi.org/}%
	\providecommand \selectlanguage [0]{\@gobble}%
	\providecommand \bibinfo  [0]{\@secondoftwo}%
	\providecommand \bibfield  [0]{\@secondoftwo}%
	\providecommand \translation [1]{[#1]}%
	\providecommand \BibitemOpen [0]{}%
	\providecommand \bibitemStop [0]{}%
	\providecommand \bibitemNoStop [0]{.\EOS\space}%
	\providecommand \EOS [0]{\spacefactor3000\relax}%
	\providecommand \BibitemShut  [1]{\csname bibitem#1\endcsname}%
	\let\auto@bib@innerbib\@empty
	%</preamble>
	\bibitem [{\citenamefont {Casimir}\ and\ \citenamefont
		{Polder}(1948)}]{Casimir1948}%
	\BibitemOpen
	\bibfield  {author} {\bibinfo {author} {\bibfnamefont {H.~B.~G.}\
			\bibnamefont {Casimir}}\ and\ \bibinfo {author} {\bibfnamefont
			{D.}~\bibnamefont {Polder}},\ }\href {\doibase 10.1103/PhysRev.73.360}
	{\bibfield  {journal} {\bibinfo  {journal} {Phys. Rev.}\ }\textbf {\bibinfo
			{volume} {73}},\ \bibinfo {pages} {360} (\bibinfo {year} {1948})}\BibitemShut
	{NoStop}%
	\bibitem [{\citenamefont {Sukenik}\ \emph {et~al.}(1993)\citenamefont
		{Sukenik}, \citenamefont {Boshier}, \citenamefont {Cho}, \citenamefont
		{Sandoghdar},\ and\ \citenamefont {Hinds}}]{Sukenik1993}%
	\BibitemOpen
	\bibfield  {author} {\bibinfo {author} {\bibfnamefont {C.~I.}\ \bibnamefont
			{Sukenik}}, \bibinfo {author} {\bibfnamefont {M.~G.}\ \bibnamefont
			{Boshier}}, \bibinfo {author} {\bibfnamefont {D.}~\bibnamefont {Cho}},
		\bibinfo {author} {\bibfnamefont {V.}~\bibnamefont {Sandoghdar}}, \ and\
		\bibinfo {author} {\bibfnamefont {E.~A.}\ \bibnamefont {Hinds}},\ }\href
	{\doibase 10.1103/PhysRevLett.70.560} {\bibfield  {journal} {\bibinfo
			{journal} {Phys. Rev. Lett.}\ }\textbf {\bibinfo {volume} {70}},\ \bibinfo
		{pages} {560} (\bibinfo {year} {1993})}\BibitemShut {NoStop}%
	\bibitem [{\citenamefont {Lamoreaux}(1997)}]{Lamoreaux1997}%
	\BibitemOpen
	\bibfield  {author} {\bibinfo {author} {\bibfnamefont {S.~K.}\ \bibnamefont
			{Lamoreaux}},\ }\href {\doibase 10.1103/PhysRevLett.78.5} {\bibfield
		{journal} {\bibinfo  {journal} {Phys. Rev. Lett.}\ }\textbf {\bibinfo
			{volume} {78}},\ \bibinfo {pages} {5} (\bibinfo {year} {1997})}\BibitemShut
	{NoStop}%
	\bibitem [{\citenamefont {Mohideen}\ and\ \citenamefont
		{Roy}(1998)}]{Mohideen1998}%
	\BibitemOpen
	\bibfield  {author} {\bibinfo {author} {\bibfnamefont {U.}~\bibnamefont
			{Mohideen}}\ and\ \bibinfo {author} {\bibfnamefont {A.}~\bibnamefont {Roy}},\
	}\href {\doibase 10.1103/PhysRevLett.81.4549} {\bibfield  {journal} {\bibinfo
			{journal} {Phys. Rev. Lett.}\ }\textbf {\bibinfo {volume} {81}},\ \bibinfo
		{pages} {4549} (\bibinfo {year} {1998})}\BibitemShut {NoStop}%
	\bibitem [{\citenamefont {Schwinger}\ \emph {et~al.}(1978)\citenamefont
		{Schwinger}, \citenamefont {DeRaad},\ and\ \citenamefont
		{Milton}}]{Schwinger1978}%
	\BibitemOpen
	\bibfield  {author} {\bibinfo {author} {\bibfnamefont {J.}~\bibnamefont
			{Schwinger}}, \bibinfo {author} {\bibfnamefont {L.~L.}\ \bibnamefont
			{DeRaad}}, \ and\ \bibinfo {author} {\bibfnamefont {K.~A.}\ \bibnamefont
			{Milton}},\ }\href {\doibase 10.1016/0003-4916(78)90172-0} {\bibfield
		{journal} {\bibinfo  {journal} {Ann. Phys. (N. Y).}\ }\textbf {\bibinfo
			{volume} {115}},\ \bibinfo {pages} {1} (\bibinfo {year} {1978})}\BibitemShut
	{NoStop}%
	\bibitem [{\citenamefont {Balian}\ and\ \citenamefont
		{Duplantier}(1978)}]{Balian1978}%
	\BibitemOpen
	\bibfield  {author} {\bibinfo {author} {\bibfnamefont {R.}~\bibnamefont
			{Balian}}\ and\ \bibinfo {author} {\bibfnamefont {B.}~\bibnamefont
			{Duplantier}},\ }\href {\doibase 10.1016/0003-4916(78)90083-0} {\bibfield
		{journal} {\bibinfo  {journal} {Ann. Phys. (N. Y).}\ }\textbf {\bibinfo
			{volume} {112}},\ \bibinfo {pages} {165} (\bibinfo {year}
		{1978})}\BibitemShut {NoStop}%
	\bibitem [{\citenamefont {Plunien}\ \emph {et~al.}(1986)\citenamefont
		{Plunien}, \citenamefont {M{\"{u}}ller},\ and\ \citenamefont
		{Greiner}}]{Plunien1986}%
	\BibitemOpen
	\bibfield  {author} {\bibinfo {author} {\bibfnamefont {G.}~\bibnamefont
			{Plunien}}, \bibinfo {author} {\bibfnamefont {B.}~\bibnamefont
			{M{\"{u}}ller}}, \ and\ \bibinfo {author} {\bibfnamefont {W.}~\bibnamefont
			{Greiner}},\ }\href {\doibase 10.1016/0370-1573(86)90020-7} {\enquote
		{\bibinfo {title} {{The Casimir effect}},}\ } (\bibinfo {year}
	{1986})\BibitemShut {NoStop}%
	\bibitem [{\citenamefont {Bordag}\ \emph {et~al.}(2001)\citenamefont {Bordag},
		\citenamefont {Mohideen},\ and\ \citenamefont {Mostepanenko}}]{Bordag2001}%
	\BibitemOpen
	\bibfield  {author} {\bibinfo {author} {\bibfnamefont {M.}~\bibnamefont
			{Bordag}}, \bibinfo {author} {\bibfnamefont {U.}~\bibnamefont {Mohideen}}, \
		and\ \bibinfo {author} {\bibfnamefont {V.~M.}\ \bibnamefont {Mostepanenko}},\
	}\href {\doibase 10.1016/S0370-1573(01)00015-1} {\enquote {\bibinfo {title}
			{{New developments in the Casimir effect}},}\ } (\bibinfo {year} {2001}),\
	\Eprint {http://arxiv.org/abs/0106045} {arXiv:0106045 [quant-ph]}
	\BibitemShut {NoStop}%
	\bibitem [{\citenamefont {Harber}\ \emph {et~al.}(2005)\citenamefont {Harber},
		\citenamefont {Obrecht}, \citenamefont {McGuirk},\ and\ \citenamefont
		{Cornell}}]{Harber2005a}%
	\BibitemOpen
	\bibfield  {author} {\bibinfo {author} {\bibfnamefont {D.~M.}\ \bibnamefont
			{Harber}}, \bibinfo {author} {\bibfnamefont {J.~M.}\ \bibnamefont {Obrecht}},
		\bibinfo {author} {\bibfnamefont {J.~M.}\ \bibnamefont {McGuirk}}, \ and\
		\bibinfo {author} {\bibfnamefont {E.~A.}\ \bibnamefont {Cornell}},\ }\href
	{\doibase 10.1103/PhysRevA.72.033610} {\bibfield  {journal} {\bibinfo
			{journal} {Phys. Rev. A - At. Mol. Opt. Phys.}\ }\textbf {\bibinfo {volume}
			{72}} (\bibinfo {year} {2005}),\ 10.1103/PhysRevA.72.033610},\ \Eprint
	{http://arxiv.org/abs/0506208} {arXiv:0506208 [cond-mat]} \BibitemShut
	{NoStop}%
	\bibitem [{\citenamefont {Obrecht}\ \emph {et~al.}(2007)\citenamefont
		{Obrecht}, \citenamefont {Wild}, \citenamefont {Antezza}, \citenamefont
		{Pitaevskii}, \citenamefont {Stringari},\ and\ \citenamefont
		{Cornell}}]{Obrecht2007}%
	\BibitemOpen
	\bibfield  {author} {\bibinfo {author} {\bibfnamefont {J.~M.}\ \bibnamefont
			{Obrecht}}, \bibinfo {author} {\bibfnamefont {R.~J.}\ \bibnamefont {Wild}},
		\bibinfo {author} {\bibfnamefont {M.}~\bibnamefont {Antezza}}, \bibinfo
		{author} {\bibfnamefont {L.~P.}\ \bibnamefont {Pitaevskii}}, \bibinfo
		{author} {\bibfnamefont {S.}~\bibnamefont {Stringari}}, \ and\ \bibinfo
		{author} {\bibfnamefont {E.~A.}\ \bibnamefont {Cornell}},\ }\href {\doibase
		10.1103/PhysRevLett.98.063201} {\bibfield  {journal} {\bibinfo  {journal}
			{Phys. Rev. Lett.}\ }\textbf {\bibinfo {volume} {98}} (\bibinfo {year}
		{2007}),\ 10.1103/PhysRevLett.98.063201},\ \Eprint
	{http://arxiv.org/abs/0608074} {arXiv:0608074 [physics]} \BibitemShut
	{NoStop}%
	\bibitem [{\citenamefont {Klimchitskaya}\ \emph {et~al.}(2009)\citenamefont
		{Klimchitskaya}, \citenamefont {Mohideen},\ and\ \citenamefont
		{Mostepanenko}}]{Klimchitskaya2009}%
	\BibitemOpen
	\bibfield  {author} {\bibinfo {author} {\bibfnamefont {G.~L.}\ \bibnamefont
			{Klimchitskaya}}, \bibinfo {author} {\bibfnamefont {U.}~\bibnamefont
			{Mohideen}}, \ and\ \bibinfo {author} {\bibfnamefont {V.~M.}\ \bibnamefont
			{Mostepanenko}},\ }\href {\doibase 10.1103/RevModPhys.81.1827} {\bibfield
		{journal} {\bibinfo  {journal} {Rev. Mod. Phys.}\ }\textbf {\bibinfo {volume}
			{81}},\ \bibinfo {pages} {1827} (\bibinfo {year} {2009})},\ \Eprint
	{http://arxiv.org/abs/0902.4022} {arXiv:0902.4022} \BibitemShut {NoStop}%
	\bibitem [{\citenamefont {Dzyaloshinskii}\ \emph {et~al.}(1961)\citenamefont
		{Dzyaloshinskii}, \citenamefont {Lifshitz},\ and\ \citenamefont
		{Pitaevskii}}]{Dzyaloshinskii1961}%
	\BibitemOpen
	\bibfield  {author} {\bibinfo {author} {\bibfnamefont {I.~E.}\ \bibnamefont
			{Dzyaloshinskii}}, \bibinfo {author} {\bibfnamefont {E.~M.}\ \bibnamefont
			{Lifshitz}}, \ and\ \bibinfo {author} {\bibfnamefont {L.~P.}\ \bibnamefont
			{Pitaevskii}},\ }\href {\doibase 10.1070/pu1961v004n02abeh003330} {\bibfield
		{journal} {\bibinfo  {journal} {Sov. Phys. Uspekhi}\ }\textbf {\bibinfo
			{volume} {4}},\ \bibinfo {pages} {153} (\bibinfo {year} {1961})}\BibitemShut
	{NoStop}%
	\bibitem [{\citenamefont {Boyer}(1974)}]{Boyer1974}%
	\BibitemOpen
	\bibfield  {author} {\bibinfo {author} {\bibfnamefont {T.~H.}\ \bibnamefont
			{Boyer}},\ }\href {\doibase 10.1103/PhysRevA.9.2078} {\bibfield  {journal}
		{\bibinfo  {journal} {Phys. Rev. A}\ }\textbf {\bibinfo {volume} {9}},\
		\bibinfo {pages} {2078} (\bibinfo {year} {1974})}\BibitemShut {NoStop}%
	\bibitem [{\citenamefont {Kenneth}\ \emph {et~al.}(2002)\citenamefont
		{Kenneth}, \citenamefont {Klich}, \citenamefont {Mann},\ and\ \citenamefont
		{Revzen}}]{Kenneth2002}%
	\BibitemOpen
	\bibfield  {author} {\bibinfo {author} {\bibfnamefont {O.}~\bibnamefont
			{Kenneth}}, \bibinfo {author} {\bibfnamefont {I.}~\bibnamefont {Klich}},
		\bibinfo {author} {\bibfnamefont {A.}~\bibnamefont {Mann}}, \ and\ \bibinfo
		{author} {\bibfnamefont {M.}~\bibnamefont {Revzen}},\ }\href {\doibase
		10.1103/PhysRevLett.89.033001} {\bibfield  {journal} {\bibinfo  {journal}
			{Phys. Rev. Lett.}\ }\textbf {\bibinfo {volume} {89}} (\bibinfo {year}
		{2002}),\ 10.1103/PhysRevLett.89.033001},\ \Eprint
	{http://arxiv.org/abs/0202114} {arXiv:0202114 [quant-ph]} \BibitemShut
	{NoStop}%
	\bibitem [{\citenamefont {Levin}\ \emph {et~al.}(2010)\citenamefont {Levin},
		\citenamefont {McCauley}, \citenamefont {Rodriguez}, \citenamefont {Reid},\
		and\ \citenamefont {Johnson}}]{Levin2010}%
	\BibitemOpen
	\bibfield  {author} {\bibinfo {author} {\bibfnamefont {M.}~\bibnamefont
			{Levin}}, \bibinfo {author} {\bibfnamefont {A.~P.}\ \bibnamefont {McCauley}},
		\bibinfo {author} {\bibfnamefont {A.~W.}\ \bibnamefont {Rodriguez}}, \bibinfo
		{author} {\bibfnamefont {M.~T.}\ \bibnamefont {Reid}}, \ and\ \bibinfo
		{author} {\bibfnamefont {S.~G.}\ \bibnamefont {Johnson}},\ }\href {\doibase
		10.1103/PhysRevLett.105.090403} {\bibfield  {journal} {\bibinfo  {journal}
			{Phys. Rev. Lett.}\ }\textbf {\bibinfo {volume} {105}} (\bibinfo {year}
		{2010}),\ 10.1103/PhysRevLett.105.090403},\ \Eprint
	{http://arxiv.org/abs/1003.3487} {arXiv:1003.3487} \BibitemShut {NoStop}%
	\bibitem [{\citenamefont {Munday}\ \emph {et~al.}(2009)\citenamefont {Munday},
		\citenamefont {Capasso},\ and\ \citenamefont {Parsegian}}]{Munday2009}%
	\BibitemOpen
	\bibfield  {author} {\bibinfo {author} {\bibfnamefont {J.~N.}\ \bibnamefont
			{Munday}}, \bibinfo {author} {\bibfnamefont {F.}~\bibnamefont {Capasso}}, \
		and\ \bibinfo {author} {\bibfnamefont {V.~A.}\ \bibnamefont {Parsegian}},\
	}\href {\doibase 10.1038/nature07610} {\bibfield  {journal} {\bibinfo
			{journal} {Nature}\ }\textbf {\bibinfo {volume} {457}},\ \bibinfo {pages}
		{170} (\bibinfo {year} {2009})}\BibitemShut {NoStop}%
	\bibitem [{\citenamefont {DeWitt}(1979)}]{DeWitt1979}%
	\BibitemOpen
	\bibfield  {author} {\bibinfo {author} {\bibfnamefont {B.~S.}\ \bibnamefont
			{DeWitt}},\ }in\ \href@noop {} {\emph {\bibinfo {booktitle} {Gen. Relativ. An
				Einstein Centen. Surv.}}},\ \bibinfo {editor} {edited by\ \bibinfo {editor}
		{\bibfnamefont {S.~W.}\ \bibnamefont {Hawking}}\ and\ \bibinfo {editor}
		{\bibfnamefont {W.}~\bibnamefont {Israel}}}\ (\bibinfo  {publisher}
	{Cambridge University Press},\ \bibinfo {address} {Cambridge},\ \bibinfo
	{year} {1979})\ pp.\ \bibinfo {pages} {680--745}\BibitemShut {NoStop}%
	\bibitem [{\citenamefont {Alhambra}\ \emph
		{et~al.}(2014{\natexlab{a}})\citenamefont {Alhambra}, \citenamefont {Kempf},\
		and\ \citenamefont {Mart{\'{i}}n-Mart{\'{i}}nez}}]{Alhambra2014}%
	\BibitemOpen
	\bibfield  {author} {\bibinfo {author} {\bibfnamefont {{\'{A}}.~M.}\
			\bibnamefont {Alhambra}}, \bibinfo {author} {\bibfnamefont {A.}~\bibnamefont
			{Kempf}}, \ and\ \bibinfo {author} {\bibfnamefont {E.}~\bibnamefont
			{Mart{\'{i}}n-Mart{\'{i}}nez}},\ }\href {\doibase 10.1103/PhysRevA.89.033835}
	{\bibfield  {journal} {\bibinfo  {journal} {Phys. Rev. A - At. Mol. Opt.
				Phys.}\ }\textbf {\bibinfo {volume} {89}} (\bibinfo {year}
		{2014}{\natexlab{a}}),\ 10.1103/PhysRevA.89.033835}\BibitemShut {NoStop}%
	\bibitem [{\citenamefont {Crispino}\ \emph {et~al.}(2008)\citenamefont
		{Crispino}, \citenamefont {Higuchi},\ and\ \citenamefont
		{Matsas}}]{Crispino2008}%
	\BibitemOpen
	\bibfield  {author} {\bibinfo {author} {\bibfnamefont {L.~C.~B.}\
			\bibnamefont {Crispino}}, \bibinfo {author} {\bibfnamefont {A.}~\bibnamefont
			{Higuchi}}, \ and\ \bibinfo {author} {\bibfnamefont {G.~E.~A.}\ \bibnamefont
			{Matsas}},\ }\href {\doibase 10.1103/RevModPhys.80.787} {\bibfield  {journal}
		{\bibinfo  {journal} {Rev. Mod. Phys.}\ }\textbf {\bibinfo {volume} {80}},\
		\bibinfo {pages} {787} (\bibinfo {year} {2008})},\ \Eprint
	{http://arxiv.org/abs/0710.5373} {arXiv:0710.5373} \BibitemShut {NoStop}%
	\bibitem [{\citenamefont {Birrell}\ and\ \citenamefont
		{Davies}(1982)}]{Birrell1982}%
	\BibitemOpen
	\bibfield  {author} {\bibinfo {author} {\bibfnamefont {N.~D.}\ \bibnamefont
			{Birrell}}\ and\ \bibinfo {author} {\bibfnamefont {P.~C.~W.}\ \bibnamefont
			{Davies}},\ }\href
	{https://books.google.pl/books/about/Quantum{\_}Fields{\_}in{\_}Curved{\_}Space.html?id=SEnaUnrqzrUC{\&}redir{\_}esc=y}
	{\emph {\bibinfo {title} {{Quantum fields in curved space}}}}\ (\bibinfo
	{publisher} {Cambridge University Press},\ \bibinfo {year} {1982})\ p.\
	\bibinfo {pages} {340}\BibitemShut {NoStop}%
	\bibitem [{\citenamefont {Louko}\ and\ \citenamefont {Satz}(2008)}]{Louko2008}%
	\BibitemOpen
	\bibfield  {author} {\bibinfo {author} {\bibfnamefont {J.}~\bibnamefont
			{Louko}}\ and\ \bibinfo {author} {\bibfnamefont {A.}~\bibnamefont {Satz}},\
	}\href {\doibase 10.1088/0264-9381/25/5/055012} {\bibfield  {journal}
		{\bibinfo  {journal} {Class. Quantum Gravity}\ }\textbf {\bibinfo {volume}
			{25}} (\bibinfo {year} {2008}),\ 10.1088/0264-9381/25/5/055012},\ \Eprint
	{http://arxiv.org/abs/0710.5671} {arXiv:0710.5671} \BibitemShut {NoStop}%
	\bibitem [{\citenamefont {Hodgkinson}(2013)}]{Hodgkinson2013}%
	\BibitemOpen
	\bibfield  {author} {\bibinfo {author} {\bibfnamefont {L.}~\bibnamefont
			{Hodgkinson}},\ }\emph {\bibinfo {title} {{Particle detectors in curved
				spacetime quantum field theory}}},\ \href@noop {} {Ph.D. thesis} (\bibinfo
	{year} {2013}),\ \Eprint {http://arxiv.org/abs/1309.7281v2}
	{arXiv:1309.7281v2} \BibitemShut {NoStop}%
	\bibitem [{\citenamefont {Brown}\ \emph {et~al.}(2013)\citenamefont {Brown},
		\citenamefont {Mart{\'{i}}n-Mart{\'{i}}nez}, \citenamefont {Menicucci},\ and\
		\citenamefont {Mann}}]{Brown2013}%
	\BibitemOpen
	\bibfield  {author} {\bibinfo {author} {\bibfnamefont {E.~G.}\ \bibnamefont
			{Brown}}, \bibinfo {author} {\bibfnamefont {E.}~\bibnamefont
			{Mart{\'{i}}n-Mart{\'{i}}nez}}, \bibinfo {author} {\bibfnamefont {N.~C.}\
			\bibnamefont {Menicucci}}, \ and\ \bibinfo {author} {\bibfnamefont {R.~B.}\
			\bibnamefont {Mann}},\ }\href {\doibase 10.1103/PhysRevD.87.084062}
	{\bibfield  {journal} {\bibinfo  {journal} {Phys. Rev. D - Part. Fields,
				Gravit. Cosmol.}\ }\textbf {\bibinfo {volume} {87}} (\bibinfo {year}
		{2013}),\ 10.1103/PhysRevD.87.084062}\BibitemShut {NoStop}%
	\bibitem [{\citenamefont {Mart{\'{i}}n-Mart{\'{i}}nez}\ \emph
		{et~al.}(2013)\citenamefont {Mart{\'{i}}n-Mart{\'{i}}nez}, \citenamefont
		{Montero},\ and\ \citenamefont {{Del Rey}}}]{Martin-Martinez2013a}%
	\BibitemOpen
	\bibfield  {author} {\bibinfo {author} {\bibfnamefont {E.}~\bibnamefont
			{Mart{\'{i}}n-Mart{\'{i}}nez}}, \bibinfo {author} {\bibfnamefont
			{M.}~\bibnamefont {Montero}}, \ and\ \bibinfo {author} {\bibfnamefont
			{M.}~\bibnamefont {{Del Rey}}},\ }\href {\doibase 10.1103/PhysRevD.87.064038}
	{\bibfield  {journal} {\bibinfo  {journal} {Phys. Rev. D - Part. Fields,
				Gravit. Cosmol.}\ }\textbf {\bibinfo {volume} {87}} (\bibinfo {year}
		{2013}),\ 10.1103/PhysRevD.87.064038},\ \Eprint
	{http://arxiv.org/abs/1207.3248} {arXiv:1207.3248} \BibitemShut {NoStop}%
	\bibitem [{\citenamefont {Passante}\ \emph {et~al.}(2003)\citenamefont
		{Passante}, \citenamefont {Persico},\ and\ \citenamefont
		{Rizzuto}}]{Passante2003}%
	\BibitemOpen
	\bibfield  {author} {\bibinfo {author} {\bibfnamefont {R.}~\bibnamefont
			{Passante}}, \bibinfo {author} {\bibfnamefont {F.}~\bibnamefont {Persico}}, \
		and\ \bibinfo {author} {\bibfnamefont {L.}~\bibnamefont {Rizzuto}},\ }\href
	{\doibase 10.1016/S0375-9601(03)01131-9} {\bibfield  {journal} {\bibinfo
			{journal} {Phys. Lett. Sect. A Gen. At. Solid State Phys.}\ }\textbf
		{\bibinfo {volume} {316}},\ \bibinfo {pages} {29} (\bibinfo {year}
		{2003})}\BibitemShut {NoStop}%
	\bibitem [{\citenamefont {Rizzuto}\ \emph {et~al.}(2004)\citenamefont
		{Rizzuto}, \citenamefont {Passante},\ and\ \citenamefont
		{Persico}}]{Rizzuto2004}%
	\BibitemOpen
	\bibfield  {author} {\bibinfo {author} {\bibfnamefont {L.}~\bibnamefont
			{Rizzuto}}, \bibinfo {author} {\bibfnamefont {R.}~\bibnamefont {Passante}}, \
		and\ \bibinfo {author} {\bibfnamefont {F.}~\bibnamefont {Persico}},\ }\href
	{\doibase 10.1103/PhysRevA.70.012107} {\enquote {\bibinfo {title} {{Dynamical
					casimir-polder energy between an excited- and a ground-state atom}},}\ }
	(\bibinfo {year} {2004})\BibitemShut {NoStop}%
	\bibitem [{\citenamefont {Spagnolo}\ \emph {et~al.}(2006)\citenamefont
		{Spagnolo}, \citenamefont {Passante},\ and\ \citenamefont
		{Rizzuto}}]{Spagnolo2006}%
	\BibitemOpen
	\bibfield  {author} {\bibinfo {author} {\bibfnamefont {S.}~\bibnamefont
			{Spagnolo}}, \bibinfo {author} {\bibfnamefont {R.}~\bibnamefont {Passante}},
		\ and\ \bibinfo {author} {\bibfnamefont {L.}~\bibnamefont {Rizzuto}},\ }\href
	{\doibase 10.1103/PhysRevA.73.062117} {\bibfield  {journal} {\bibinfo
			{journal} {Phys. Rev. A - At. Mol. Opt. Phys.}\ }\textbf {\bibinfo {volume}
			{73}} (\bibinfo {year} {2006}),\ 10.1103/PhysRevA.73.062117}\BibitemShut
	{NoStop}%
	\bibitem [{\citenamefont {Rizzuto}(2007)}]{Rizzuto2007}%
	\BibitemOpen
	\bibfield  {author} {\bibinfo {author} {\bibfnamefont {L.}~\bibnamefont
			{Rizzuto}},\ }\href {\doibase 10.1103/PhysRevA.76.062114} {\bibfield
		{journal} {\bibinfo  {journal} {Phys. Rev. A - At. Mol. Opt. Phys.}\ }\textbf
		{\bibinfo {volume} {76}} (\bibinfo {year} {2007}),\
		10.1103/PhysRevA.76.062114}\BibitemShut {NoStop}%
	\bibitem [{\citenamefont {Alhambra}\ \emph
		{et~al.}(2014{\natexlab{b}})\citenamefont {Alhambra}, \citenamefont {Kempf},\
		and\ \citenamefont {Mart{\'{i}}n-Mart{\'{i}}nez}}]{Alhambra2014a}%
	\BibitemOpen
	\bibfield  {author} {\bibinfo {author} {\bibfnamefont {{\'{A}}.~M.}\
			\bibnamefont {Alhambra}}, \bibinfo {author} {\bibfnamefont {A.}~\bibnamefont
			{Kempf}}, \ and\ \bibinfo {author} {\bibfnamefont {E.}~\bibnamefont
			{Mart{\'{i}}n-Mart{\'{i}}nez}},\ }\href {\doibase 10.1103/PhysRevA.89.033835}
	{\bibfield  {journal} {\bibinfo  {journal} {Phys. Rev. A - At. Mol. Opt.
				Phys.}\ }\textbf {\bibinfo {volume} {89}} (\bibinfo {year}
		{2014}{\natexlab{b}}),\ 10.1103/PhysRevA.89.033835}\BibitemShut {NoStop}%
\end{thebibliography}

	\appendix
	\section{Low frequency approximation}
	In this Appendix we demonstrate that the infinite series given by the \eqref{eqn:ECP} and \eqref{eqn:pav}, can be well approximated by a sum of finite number of terms.
	We will include only the lowest modes of the field to the sum, so it can be called a low frequency approximation.
	Furthermore, we want to show that the frequency of the last mode included in the approximation sum always stays much smaller than the energy gap $\Omega$.
	
		\begin{figure}[h!tbp]
		\includegraphics[width=1\linewidth]{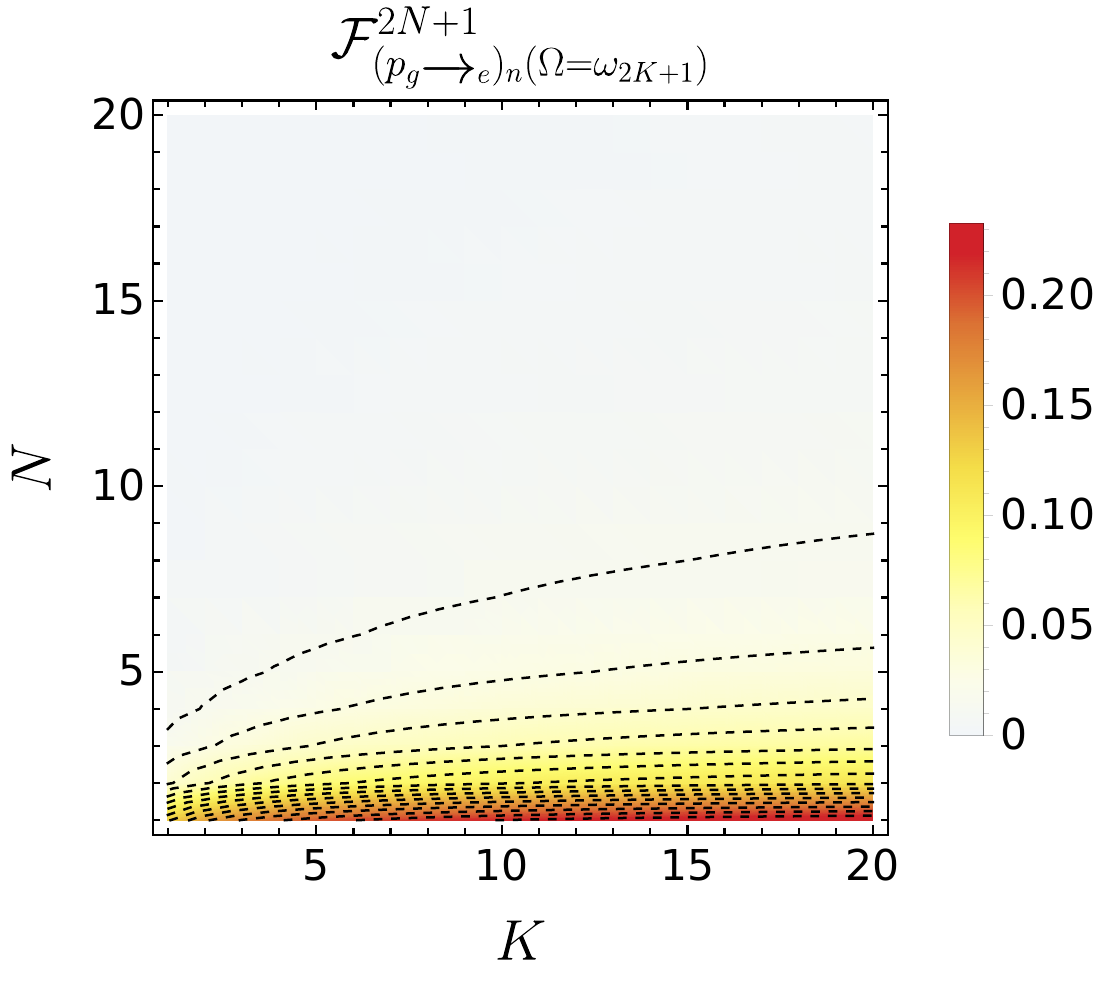}
		\includegraphics[width=1\linewidth]{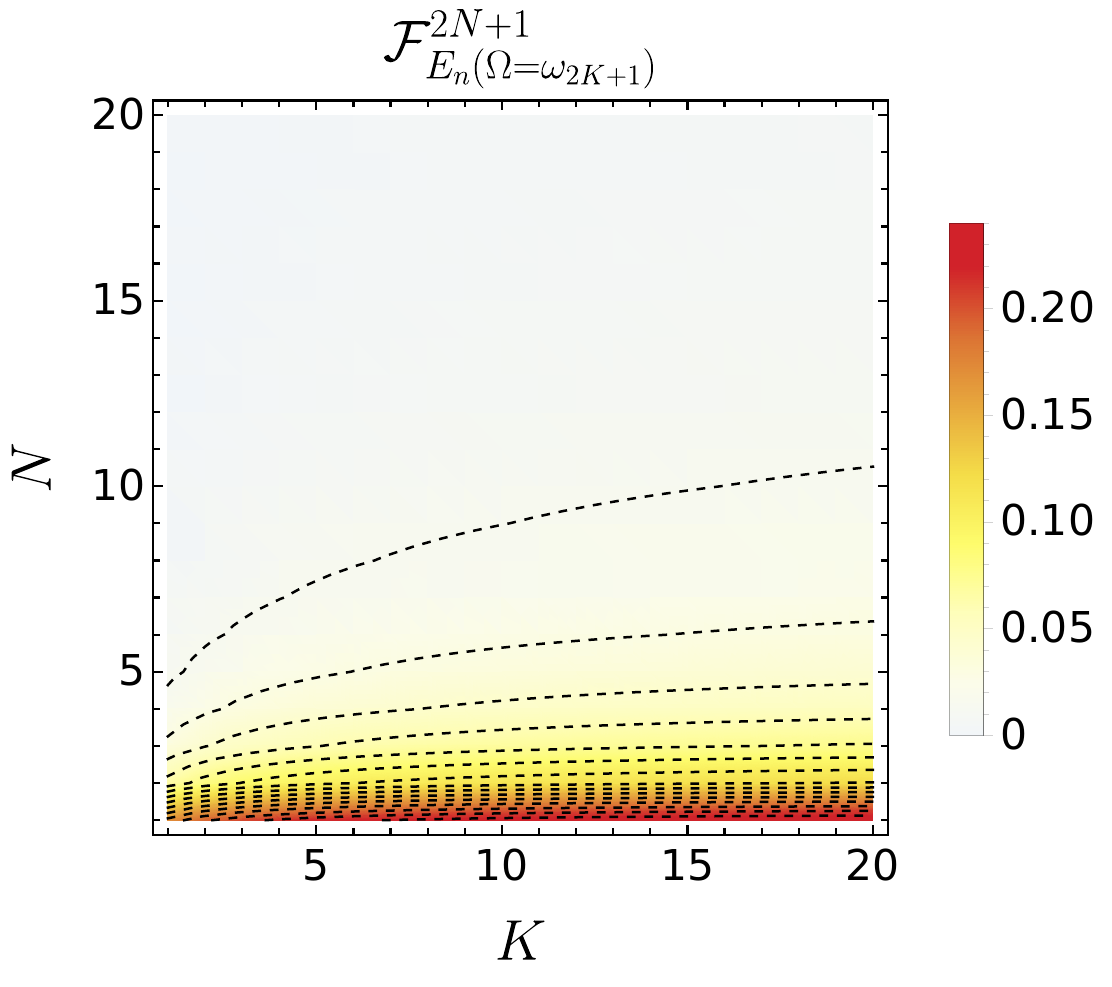}
		\includegraphics[width=1\linewidth]{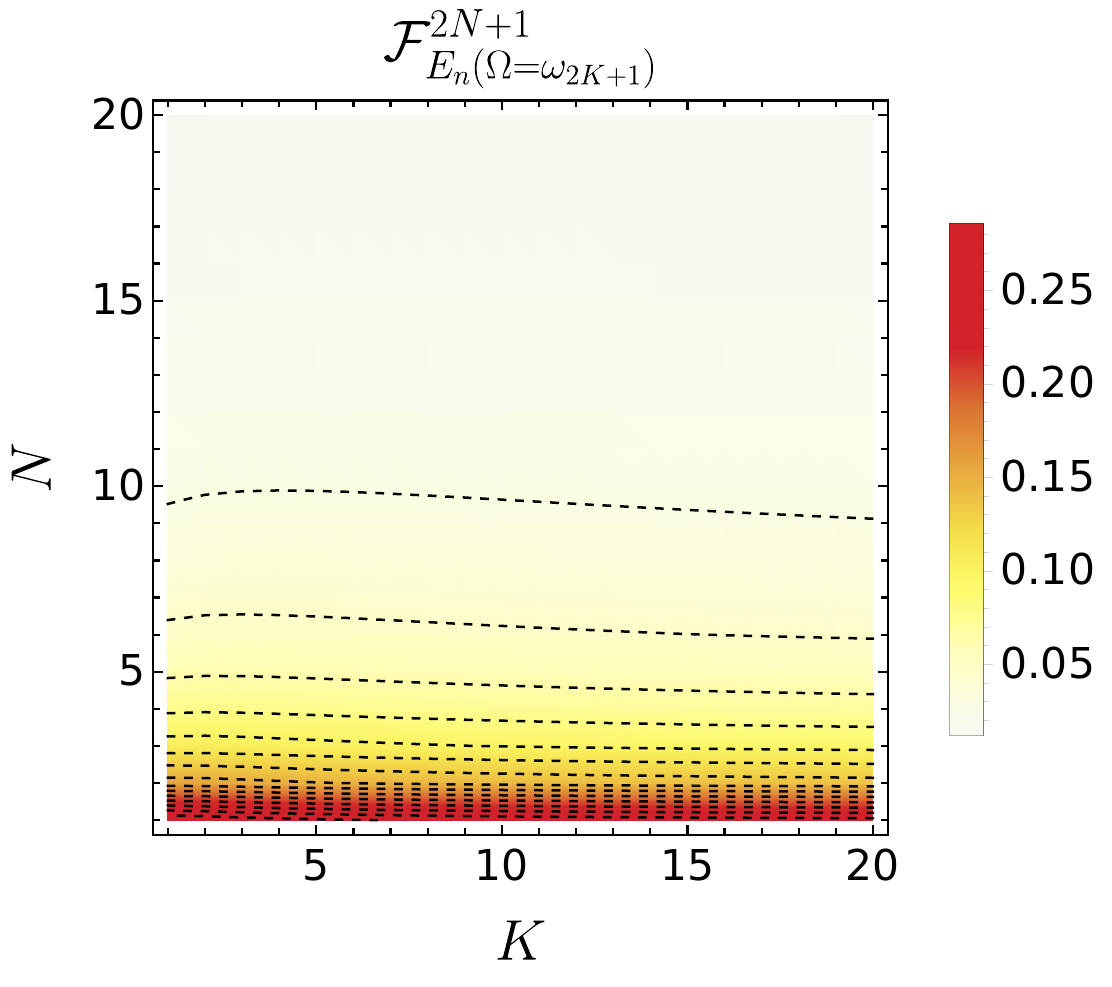}
		\caption{Fidelity of the finite elements approximation used to find a value of Casimir-Polder potential or probability of excitation for a detector standing in the middle of the cavity.
			The parameters taken for the plot read: $x=\frac{L}{2},~L=1,~m=1\cdot10^{-3},~\lambda=1\cdot10^{-2}$ (in the case of finding Casimir-Polder potential there is also $\alpha$ coupling constant).
			Fidelity of:
			(Top) probability of excitation,
			(Center) Casimir-Polder potential for $\alpha=1\cdot10^{-1}$,
			(Bottom) Casimir-Polder potential for $\alpha=2$.
			\label{fig3}}
	\end{figure}
	
	Let us start from the most general case.
	Let us cosider a converged infinite sum $S=\sum_{n=1}^{\infty}a_n$.
	The value of $S$ can be approximated by the $S_N=\sum_{n=1}^{N}a_n$.
	The bigger $N$ is, the better approximation of the infinite series $S$ we get.
	We can ask how many elements of the series need to be summed up to achieve a given quality of the approximation.
	To well define this problem we have to determine how to measure the quality of the approximation of the series
	One of the possibility is to use a fidelity function defined as:
	\begin{equation}\label{fid}
	\mathcal{F}_{a_n}^{N}=\frac{a_{N}}{\sum_{n=1}^{N}a_n}.
	\end{equation}
	Such a function tells us how big contribution to finite sum $S_N$ coming from the last element is.
	The smaller $\mathcal{F}_{a_n}^{N}$ is, the better the quality of $S$ approximation given by $S_N$ becomes.
	
	In the case presented above, we want to verify whether series which define Casimir-Polder potential and probability of excitation can be approximated by a sum including just $N$ elements such that $\omega_N$ is still much smaller than $\Omega$.
	Using fidelity function~\eqref{fid} we can find the value of the function $\mathcal{F}_{E_n(\Omega)}^{N}$, where $E_n$ is such that $E_{\mathrm{CP}}=\sum_{n=1}^{\infty}E_n$.
	To answer our question we can plot $\mathcal{F}_{E_n(\Omega=\omega_K)}^{N}$ as a function of $N$ and $K$ such that $\omega_K=\Omega$.
	Similarly, for the probability of excitation we will plot $\mathcal{F}_{(p_{g\xrightarrow{}e})_n(\Omega)}^{N}$, where $(p_{g\xrightarrow{}e})_n$ is such that $p_{g\xrightarrow{}e}=\sum_{n}(p_{g\xrightarrow{}e})_n$. 
	
	For simplicity, we will consider only a detector standing in the middle of the cavity.
	As a result, only odd modes of the field have non zero contribution to the final value.
	The figure \ref{fig3} shows fidelity of the approximated series-defined Casimir-Polder potential and probability of excitation.
	We can see that the lines connecting the points of the same value of the fidelity function have a convex shape.
	
	It turns out that for every parameter describing the quality of the approximation by a $N$ elements sum, we can choose $\Omega$ for which $\omega_N\ll\Omega$.
	
	For instance, let us consider the line of constant value of fidelity shown on the top of figure \ref{fig3}.
	The same value of a fidelity occurs for pair $(N,K)\approx(7,7)$ and for $(N,K)\approx(12,20)$.
	It means that for $\Omega=\omega_{2\cdot7+1}$ one has to sum $N=2\cdot7+1$ modes of the field to achieve the same fidelity as for $\Omega=\omega_{2\cdot20+1}$ and only $N=2\cdot12+1$ modes.
	
	For series-defined Casimir-Polder potential and coupling $\alpha>1$ it is even better, because lines of constant fidelity decrease with $K$, so the bigger $\Omega=\omega_{2K+1}$ is, the smaller number of modes that are needed to be summed up to achieve given fidelity is.
	
\end{document}